# UVC LEDs on Bulk AlN Substrates Using Silicon Nanomembranes for Efficient Hole Injection

Sang June Cho[1], Dong Liu[1], Jung-Hun Seo[1], Rafael Dalmau[2],

Kwangeun Kim[1], Jeongpil Park[1], Deyin Zhao[3], Xin Yin[4], Yei Hwan Jung[1],

In-Kyu Lee[1], Munho Kim[1], Xudong Wang[4], John D. Albrecht[5,*], Weidong Zhou[3],

Baxter Moody[2,*] and Zhenqiang Ma[1, *]

[1]Department of Electrical and Computer Engineering, University of Wisconsin-Madison, 1415 Engineering Drive, Madison, WI 53706, USA

[2]HexaTech, Inc., 991 Aviation Parkway, Morrisville, North Carolina 27560, USA

[3] Department of Electrical Engineering, University of Texas at Arlington, 500 South Cooper Street, Arlington, Texas 76019, USA

[4]Department of Materials Science and Engineering, University of Wisconsin-Madison, Madison, 1509 University Avenue, Madison, WI 53706, USA

[5]Department of Electrical and Computer Engineering, Michigan State University, 428 S. Shaw Lane, East Lansing, Michigan 48824, USA

*Emails: mazq@engr.wisc.edu, jalbrech@egr.msu.edu, bmoody@hexatechinc.com

As UV LEDs are explored at shorter wavelengths (< 280 nm) into the UVC spectral range, the crystalline quality of epitaxial AlGaN films with high Al compositions and inefficient hole injection from p-type AlGaN severely limit the LED performance and development. In this work, we report on 237 nm light emission with a record light output power of 265 μW from AlN/Al$_{0.72}$Ga$_{028}$N multiple quantum well UVC LEDs using bulk AlN substrates and p-type silicon nanomembrane contact layers for significantly improved AlGaN film quality and hole injection, respectively. No intensity degradation or efficiency droop was observed up to a current density of 245 A/cm$^2$, which is attributed to the low dislocation density within AlGaN films, the large concentration of holes from **p-Si, and efficient hole-transport to the active region. Additionally, the emission peak at 237 nm is dominant across the electroluminescence spectrum with no significant parasitic emissions observed. This study demonstrates the feasibility of using p-Si as a hole injector for UVC LEDs, which can be extended to even shorter wavelengths where hole injection from chemically doped AlGaN layers is not feasible.**

Semiconductor UVC light emitting diodes (LEDs) operating at sub-280 nm wavelengths have attracted increasing interest due to their critical applications, such as biological agent detection, decontamination, medical treatment, lithography, communication, and Raman spectroscopy.[1-6] AlGaN is the most widely used material for commercial UV LEDs because of its direct bandgap spanning the UVA, UVB, and UVC ranges[7-15]. However, as sub-280 nm wavelengths are explored, the following challenges arise. First, the high dislocation density of epitaxial AlGaN device films deposited on foreign substrates leads to high concentrations of non-radiative recombination centers. Second, realizing low-resistance metal contacts become increasingly difficult with increasing Al composition. Third, and also the most challenging that also applies to UV LEDs of longer wavelengths, is the extremely poor hole injection from p-AlGaN arising from inefficient acceptor activation due to the large ionization energy of acceptors (630 meV for AlN:Mg)[8, 16] and low hole mobility in alloys.

Planar UVC LEDs are commonly grown on highly lattice-mismatched sapphire substrates resulting in a high concentration of threading dislocations (TDs) ($>10^9$ cm$^{-3}$), which act as non-radiative recombination centers contributing to a low internal quantum efficiency (IQE)[17-19]. Nanowire-based UVC LEDs have proved to substantially reduce defects and dislocation densities during growth on Si[20, 21]. In addition, nanowire UV lasers have been demonstrated at low temperatures by employing Anderson light localization[22-24] and nanowire self-organization[25]. However, as the optical cavity is randomly formed[25], the uniformity and transmission direction of light emission are unlikely to be predicted and controlled. Additionally, the lack of light power scalability of the nanostructured UV LEDs poses a challenge to many practical applications. As an alternative, monolayer GaN quantum wells and dots between AlN barriers have been demonstrated to be able to improve the internal quantum efficiency (IQE) as carriers are kept away from non-radiative recombination centers due to three-dimensional confinement[2-5, 26-29]. Light emission at 232 nm was recently reported from using 1–2 ML GaN quantum structures[5]. While ultrathin GaN wells are required to achieve the large blue-shift to UVC wavelengths, the LED output power, which was not reported[5], has to overcome the absorption losses to the n- and p- AlGaN graded (Al: 0.5 to 1) regions with a lower bandgap energy than the 232 nm photons and will ultimately limit power scaling and efficiency. In addition, due to the use of the thinnest possible GaN and the thus the fluctuations in the actual GaN thickness, achieving electroluminescence at even shorter emission wavelengths will be unlikely feasible[30].

To improve the free carrier hole concentration in high Al content AlGaN materials, researchers have developed the approach of polarization doping[1, 5, 31-33] for the purpose of enhancing electrical conductivity. By linearly grading Al composition and the corresponding electric field to take advantage of the intrinsic spontaneous polarization effect, increased carrier concentration[34-38] has been realized. However, for the approach to produce substantial effective doping, there must be a strong composition gradient (a large range of Al variation). This capability diminishes as the Al composition in AlGaN quantum wells approaches the binary endpoint, AlN, which is necessary to generate high energy photons (~5 eV for 250 nm wavelength).

In this work, concerns of non-radiative recombination due to the crystal quality are essentially eliminated by reducing the dislocation density by several orders of magnitude. Here, bulk AlN substrates are adopted to grow Al-face high Al composition $Al_xGa_{1-x}N$ epitaxial heterostructures. Epitaxial device layers inherit the low dislocation density ($<10^4$ cm$^{-3}$) of the single-crystalline bulk substrate. Consequently, efficiency degradation induced by dislocation recombination centers is no longer significant. The high quality of AlGaN epitaxial films for DUV LEDs was verified using X-ray diffraction (XRD).

To tackle the problems of poor hole injection and poor p-side conductivity, a heavily p-type doped single crystalline Si nanomembrane (Si NM) was bonded to the i-AlN/$Al_{0.72}Ga_{0.28}N$ MQW planar epi-grown structure. At the completion of Si NM bonding, the bonded device layers were characterized again using XRD. UV LEDs were fabricated using mesa etching processes. Device performance was characterized by electroluminescence (EL) measurements, including current density−voltage characteristics and EL spectrum intensity as a function of current density. No degradation of emitted light intensity was observed up to a current density of 245 A/cm$^2$, for the peak emission wavelength of 237 nm. The efficiency droop-free behavior is attributed to the large concentration of holes supplied by the p-type Si NM. The peak at 237 nm is the dominant emission over the EL spectrum without significant parasitic emission, which is common in UV LEDs and degrades efficiency. A record light output power of 265 μW was measured. Furthermore, the enhanced hole injection mechanism is explained below through a proposed band alignment arising from an in-depth examination of the hole transport across the p-Si/p-GaN heterojunction.

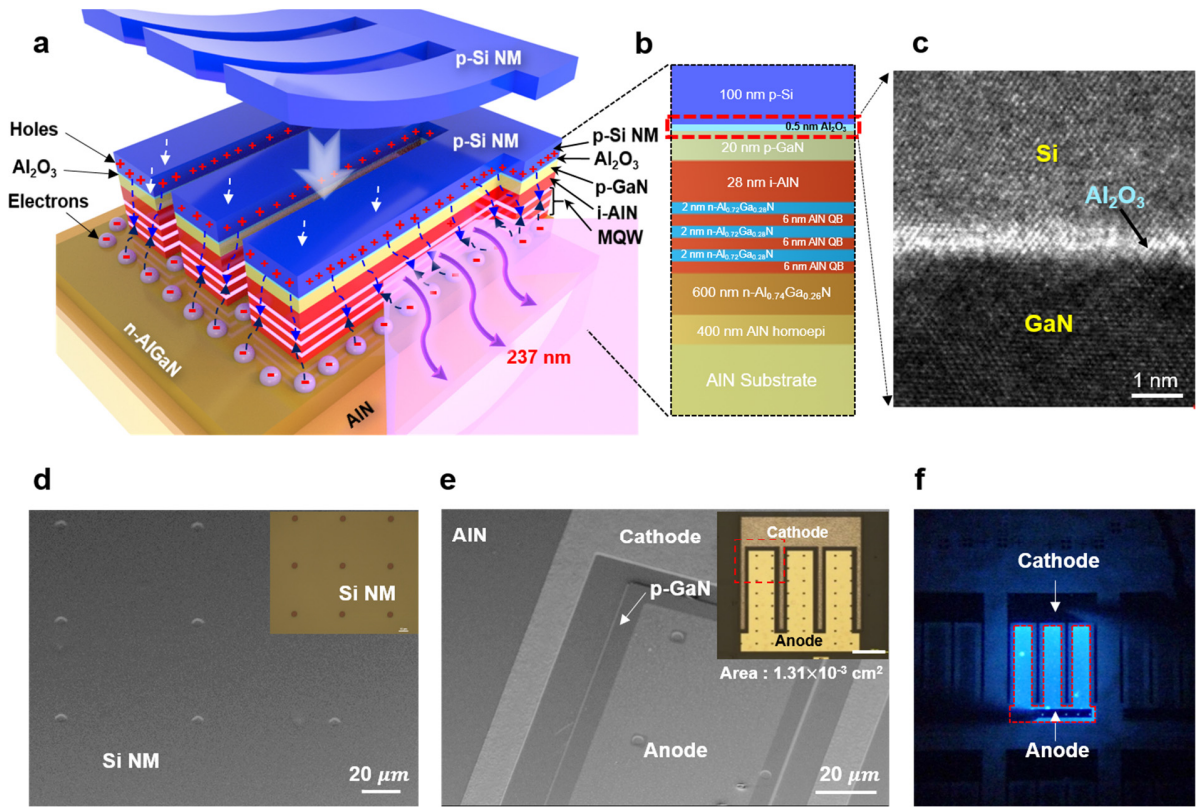

**Figure 1. UV LED schematic with Si NM hole injector, layer structure, TEM, and SEM images. a,** Three-dimensional illustration of an LED emitting UV light with a p-type single-crystal Si NM transfer process. The anode metal placed on top of the p-Si NM and the cathode metal placed on top of n-Al$_{0.74}$Ga$_{0.26}$N are not present in this illustration. **b,** UV LED epitaxial structure and p-Si/Al$_2$O$_3$/p-GaN heterojunction structure. **c,** HRTEM image of p-Si/Al$_2$O$_3$/p-GaN interface. The 0.5 nm Al$_2$O$_3$ was broadened and partially re-crystallized. **d,** An SEM image of a bonded p-Si/Al$_2$O$_3$/p-GaN structure: Si NM is transferred to Al$_2$O$_3$ deposited on an UV LED structure. Inset: an optical microscope image of a piece of Si NM. **e,** An SEM image, corresponding to the red dashed line area of the inset optical microscopic image, of a fabricated UV LED with one anode finger between two cathode fingers. Inset: an optical microscope image of the UV LED. The scale bar is 20 μm and the device area is 1.31×10$^{-3}$ cm$^2$. **f,** An optical microscope image of the UV LED under bias showing weak blue electroluminescence.

The three-dimensional UV LED structure is illustrated in Fig. 1a along with the detailed layer structure shown in Fig. 1b. The structure was grown on an AlN substrate by low pressure organometallic vapor phase epitaxy (LP-OMVPE) in a custom high-temperature reactor. The Al, Ga, N, Si, and Mg precursors are trimethylaluminum, triethylgallium, ammonia, silane, and bis(cyclopentadienyl)-magnesium, respectively, in a hydrogen diluent. As shown in Fig. 1b, following an initial 400 nm AlN homoepitaxial layer, the epitaxial portion of the active device is comprised of a Si doped 600 nm n-$Al_{0.74}Ga_{0.26}N$ electron injection layer, a 3-period 2 nm $Al_{0.72}Ga_{0.28}N$/6 nm AlN MQW, and a 28 nm AlN electron blocking layer (EBL).

Following the AlN EBL, a 20 nm Mg doped p-GaN layer was grown to circumvent the rapid oxidation of the AlN surface (see Fig. S1). On top of the epitaxial layer, a 0.5 nm $Al_2O_3$ layer was deposited by atomic layer deposition (ALD) after cleaning by the standard RCA method. The 0.5 nm $Al_2O_3$ layer acts as a quantum tunnel barrier and also a passivation layer to form the needed lattice-mismatched heterojunctions[39]. A p-type doped ($5 \times 10^{19}$ $cm^{-3}$) single-crystalline Si NM, which was pre-released from an SOI substrate, was then transferred ((Fig. S 2a (iii) and b) to the top of the $Al_2O_3$ layer, followed by a rapid thermal anneal (RTA) at 500°C for 5 minutes. The RTA procedure was intended to activate the chemical bonding process and to increase the bonding strength between the p-Si NM and $Al_2O_3$. The entire device fabrication process flow and related optical images are shown in Fig. S2.

A high-resolution transmission electron microscope (HRTEM) image of the interface between Si and GaN is shown in Fig. 1c. The HRTEM image shows that an interfacial layer of 0.46-0.99 nm with an average thickness of 0.70 nm was formed. Since ALD deposition can accurately control the thickness of the $Al_2O_3$, the slightly increased interface thickness indicates possible variations in the $Al_2O_3$. The thicker $Al_2O_3$ in some locations is presumably attributed to the chemical reaction and/or possible diffusion or reflow of $Al_2O_3$ into p-GaN and/or p-Si during the RTA annealing procedure. From the HRTEM, it can also be seen that the $Al_2O_3$ interfacial layer was partially re-crystallized with GaN and/or Si. A detailed study is needed to understand the re-crystallization mechanism. Figure 1d shows a scanning electron microscope (SEM) image of a piece of Si NM that was bonded to the epitaxial MQW layer. Figure 1e shows an SEM image of a part of a fabricated UV LED and an optical microscopic image of the entire device is shown in the inset of Fig. 1e. Figure 1f shows a 237 nm LED under forward bias with weak blue emission. The weak blue emission is most likely from the p-GaN and deep levels elsewhere in the device, and is commonly seen in the UV LED literature.

To confirm efficient hole injection from p-Si, the electrical characteristics of a p-Si/$Al_2O_3$/p-GaN isotype heterojunction (Fig. 2a and 2b) were examined. Note that in this testing structure 200 nm p-GaN was used in contrast to 20 nm GaN (only for preventing from oxidation of AlN) used in the UV LEDs. The current-voltage curves that were measured from the p-Si/$Al_2O_3$/p-GaN isotype heterojunction were plotted in the linear scale as shown in Fig. 2c. The curves indicate nearly Ohmic behavior with a turn-on voltage as small as 0.3 V. As the spacing between the metal contacts (Fig. 2a and 2b) increased from 8 μm to 10 μm and to 20 μm, the current decreases for a given voltage, as expected, due to larger lateral current spreading resistance in the p-GaN region. Furthermore, the capacitance-voltage (C-V) measurement of the p-Si/$Al_2O_3$/p-GaN heterojunction was carried out at a frequency of 1 MHz. The flat-band voltage was extracted to be -0.97 V from the $1/C^2$-V plots (Fig. 2d). The C-V measurements indicate the built-in potential $\Phi$ of -0.97 eV for the p-Si/$Al_2O_3$/p-GaN heterojunction. Since the doping concentration of p-Si ($5\times10^{19}$ cm$^{-3}$) is more than one order of magnitude higher than that of p-GaN ($1\times10^{18}$ cm$^{-3}$), it is reasonable to assume that the 0.97 eV built-in potential mainly originated from the p-GaN (one-sided junction), which was manifested with 0.97 eV downward band bending of the Ga-polar surface in the p-GaN as shown in Fig. 2e. Additionally, X-ray photoelectron spectroscopy (XPS) was employed to examine the surface/interface of p-GaN with Al$_2$O$_3$ ALD deposition (Fig. S3), which revealed a downward band bending value of 1.00 eV, roughly consistent with the 0.97 eV band bending from the C-V analysis.

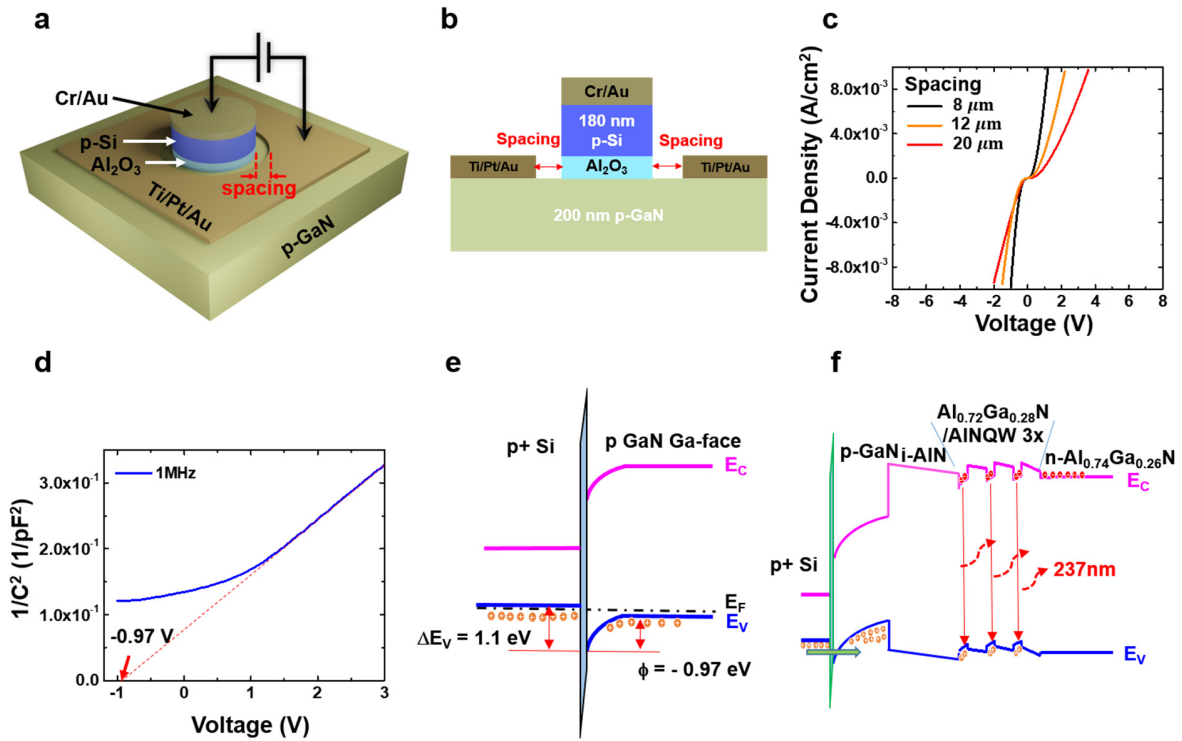

**Figure 2. Schematic of a p-Si/Al$_2$O$_3$/p-GaN isotype heterojunction, I-V, 1/C$^2$ results, and band alignment. a,** Schematics of p-Si/Al$_2$O$_3$/p-GaN isotype heterojunction test structure. The deposited thickness of Al$_2$O$_3$ by ALD is 0.5 nm. The p-GaN thickness for this testing structure is 200 nm. **b,** Measured electrical characteristics of the p-Si/Al$_2$O$_3$/p-GaN heterojunction with different metal contact distances (gap). **c,** Measured 1/C$^2$ versus voltage with a 8 μm gap distance between metal contacts at 1 MHz. **d,** Proposed band alignment of the p-Si/Al$_2$O$_3$/p-GaN under equilibrium that is responsible for the improved hole injection for the UV LEDs of this work. The charge symbols were drawn for holes.

Based on the measurements of the built-in potential, the valence band offset between p-Si and p-GaN can be determined by the following[40, 41]:

$$\Phi = \Delta E_v - \delta_{p-Si} + \delta_{p-GaN}$$

where $\Phi$ is the built-in potential, extracted as -0.97 eV. The energy differences, $\delta_{p-Si}$ and $\delta_{p-GaN}$, between Fermi levels and valence bands ($E_F$-$E_V$) for heavily doped p-Si and p-GaN (assuming an acceptor activation ratio of 10%) were calculated as -0.002 eV and 0.12 eV, respectively. Therefore, the valance band offset $|\Delta E_v|$ is 1.1 eV for the p-Si/Al$_2$O$_3$/p-GaN heterojunction. In comparison to the intrinsic $|\Delta E_v|$ value of 2.32 eV, which is obtained from the electron affinity rule, the energy barrier for hole transport from p-Si to p-GaN was substantially lowered. As a result, a much larger concentration of holes can be injected from Si into the MQW. The energy barrier difference reflected from the difference of $|\Delta E_v|$ value stems from the negative polarization charges on the surface of the Ga-face GaN as well as a voltage drop across the interfacial layer induced by the polarization charge. It is note that it is the use of 0.5 Al$_2$O$_3$ interfacial layer, which served as an effective passivation layer, that caused the reduced band downward bending of Ga-face GaN (Fig. S3 c) and thus the reduced the barrier height for hole transport.

Based on the above analyses of the surface band bending and interface-induced valance band offset shift, the band alignment of the p-Si/Al$_2$O$_3$/p-GaN isotype heterojunction under equilibrium is depicted in Fig. 2e. Under forward bias, the energy barrier for hole transport from p-Si to p-GaN was further lowered by externally applied potential to allow holes to flow, which thus enabled efficient hole transport as indicated by the electrical measurement results (Fig. 2b). The band alignment for the entire LED structure consisting of p-Si, p-GaN, i-AlN, i-AlN/Al$_{0.72}$Ga$_{0.28}$N MQWs, and n-Al$_{0.74}$Ga$_{0.26}$N contact layer under forward bias is sketched in Fig. 2f. In the LED structure, the 20 nm p-GaN should be fully depleted due to its thinness and its band bending (Fig. S2 c). Abundant holes from the p-type Si tunnel through the thin Al$_2$O$_3$ layer into p-GaN. Hole accumulation occurs at the interface adjacent to the p-AlN due to band bending from the combined effects of the built-in potential and forward bias applied, which facilitates hole transport. In this LED structure, it is the p-Si not the p-GaN that supplied holes. As a result, the p-Si serves as the actual hole injector, besides as a p-type contact layer. While the p-GaN was intended to protect the underneath AlN from oxidation, it also served as a transition layer for hole transport. Eventually, the injected holes recombine with electrons injected by the n-Al$_{0.74}$Ga$_{0.26}$N within the Al$_{0.72}$Ga$_{0.28}$N/AlN quantum wells for light emission at 237 nm. It is worth mentioning that with forward bias the valence band discontinuity $|\Delta E_v|$ could be further reduced, as voltage drop across the interfacial Al$_2$O$_3$ layer increases under forward bias.

Significantly, the p-Si hole injector approach used in this work is readily applicable to UV LEDs of any wavelengths. As the UV wavelengths become shorter, p-type doping becomes even more inefficient in the related nitride materials. As a result, the novel p-Si hole injector approach would be a critical solution for UV LEDs of very short wavelengths

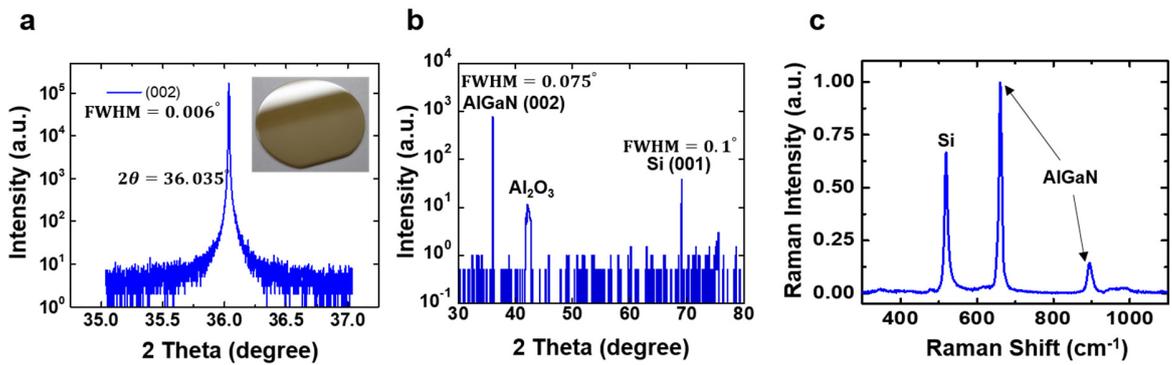

**Figure 3. XRD and Raman results of UV LED structure with Si hole injector. a,** XRD results of a bare AlN substrate. The FWHM value is 0.006 degree (21.6 arcsec). The inset shows an image of 1" diameter AlN substrate. **b,** XRD data taken from a p-Si/Al$_2$O$_3$/p-GaN/p-GaN/i-AlN/Al$_{0.72}$Ga$_{0.28}$N MQW/n-Al$_{0.74}$Ga$_{0.26}$N/AlN structure. **c,** Raman spectrum of the p-Si/Al$_2$O$_3$/p-GaN/i-AlN/Al$_{0.72}$Ga$_{0.28}$N MQW/n-Al$_{0.74}$Ga$_{0.26}$N/AlN structure.

It is crucial to maintain strain-free bonding during the bonding process, since the strain in Si NM can create unwanted band bending or even surface states at the interface. High resolution XRD (PANalytical X'pert Pro MRD) with a Cu-Kα1 radiation source (λ = 1.5406 Å) and Horiba LabRAM ARAMIS Raman confocal microscope with an 18.5 mW He-Ne (532 nm) green laser were used to assess the crystalline quality and strain status after the Si NM was bonded with a AlN/AlGaN structure. Prior to the AlN homoepitaxial layer growth, the AlN substrate as shown in the inset of Fig. 3a was examined by XRD, and indicated high crystalline quality. In Fig. 3b, the XRD spectrum taken from the p-Si/Al$_2$O$_3$/p-GaN/i-AlN/i-Al$_{0.72}$Ga$_{0.28}$N/AlN MQW/n-Al$_{0.74}$GaN$_{0.26}$/AlN structure shows three peaks corresponding to the AlGaN (002) direction, Al$_2$O$_3$, and Si (001) directions, respectively. Full width at half maximum (FWHM) values of 0.075° and 0.1° from AlGaN and Si indicate that both the AlGaN epi-layer and transferred Si

NM layer maintained good crystallinity during the transfer/bonding processes. The appearance of the weak Al$_2$O$_3$ peak at 42.1 degree in the XRD spectrum (Fig. 2b) indicates the existence of crystalline Al$_2$O$_3$, which may be attributed to recrystallization during the annealing process. The weak intensity and the wide FWHM imply that the deposited Al$_2$O$_3$ was only partially crystallized. The Al$_2$O$_3$ peak appearing in the XRD spectrum is consistent with the interface HRTEM image (Fig. 1c). The Raman spectrum shown in Fig. 3 c indicates that the Si peak and the AlGaN peaks appeared at 520.8 cm$^{-1}$ and 660.8 cm$^{-1}$/860 cm$^{-1}$, respectively, confirming that the transferred Si NM as well as AlGaN epi-layers were free of strain, which is crucial to form stable heterostructures.

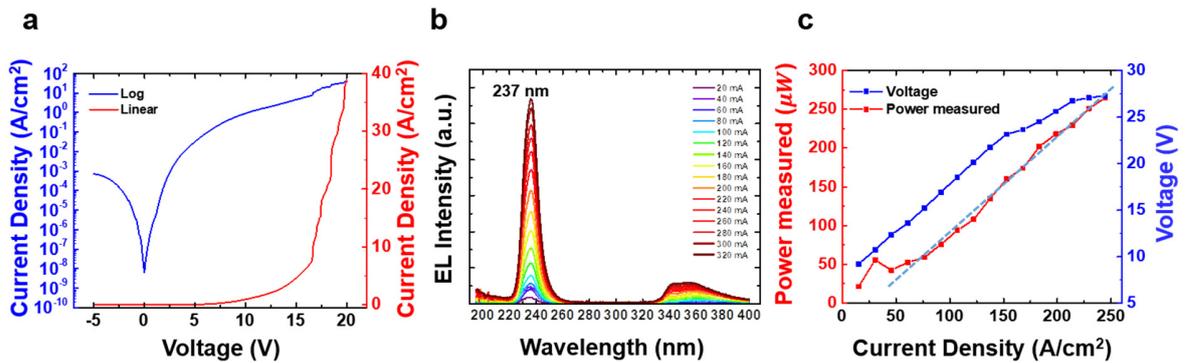

**Figure 4. Electrical characteristics and electroluminescence performance. a,** Current density-voltage characteristics of the fabricated UV LED. **b,** The linear scale of EL spectra were taken from various currents ranging from 20 mA to 320 mA. **c,** Characteristics of measured light output power and the corresponding measured voltage applied to the LED versus current density. The dotted line was drawn to show the linear trend of the light output power.

The measured current density-voltage (J-V) plot for the 237 nm LED (Fig. 4a) shows its rectifying characteristics. EL spectra and optical power measurements were performed by coupling the UV LED emission into the 6-inch integrating sphere of a Gooch and Housego OL 770-LED calibrated spectroradiometer. Electrical power was supplied in constant current mode and the temperature was not controlled, so LEDs were allowed to self-heat. The linear scale of the EL spectra for various currents from 20 mA to 320 mA are depicted in Fig. 4b, where the 237 nm emission peak from the $Al_{0.72}Ga_{0.28}N$/AlN MQWs is the dominant spectral feature. With the current ranging from 20 mA to 320 mA, the intensity of the 237 nm emission peak monotonically increases with increasing drive. Alongside the main peak emission, there is another rather weak parasitic peak detected in the near-UV range. The parasitic peak is much weaker than that of the recently reported 232 nm UV LED[5] which is likely from the top p-GaN combined with deep-levels in AlGaN, excited by 237 nm photons. The measured light output power of the LED alongside the applied voltage versus the current density (L-I-V) is plotted in Fig. 4c. It can be seen that the light output power increases linearly with current density up to 245 A/cm$^2$, which is equivalent to a current of 320 mA. This linear behavior implies an absence of efficiency droop, which directly results from the use of p-type Si as the hole injector. The large hole density, improves quantum efficiency by providing a better balance of electrons and holes than the conventional nitride-based p-type injectors[2]. Besides the efficiency droop-free behavior, 265 µW output power was measured from the UV LED at the current density of 245 A/cm$^2$ (320 mA) with an external voltage bias of 27.3 V, corresponding to an external quantum efficiency of 0.016% and a wall-plug efficiency of 0.003%. While this is not the highest efficiency[10], the result is encouraging given that reports of EL have only been demonstrated in pulsed operation mode by others up to date and our LED provides the highest UVC output power under constant current mode in comparison to all the reported output power values in literature.

In summary, the ultrathin oxide-interfaced large lattice-mismatched heterojunction approach provides to be viable toward practical implementation of DUV LEDs. Applying the approach to LEDs of even shorter wavelengths and to large LED arrays may represent the directions of future research.

# References

1. Simon, J., Protasenko, V., Lian, C., Xing, H. & Jena, D. Polarization-induced hole doping in wide-band-gap uniaxial semiconductor heterostructures. *Science* **327,** 60-64 (2010).

2. Verma, J. *et al.* Tunnel-injection GaN quantum dot ultraviolet light-emitting diodes. *Appl. Phys. Lett.* **102,** 041103 (2013).

3. Taniyasu, Y. & Kasu, M. Polarization property of deep-ultraviolet light emission from C-plane AlN/GaN short-period superlattices. *Appl. Phys. Lett.* **99,** 251112 (2011).

4. Verma, J. *et al.* Tunnel-injection quantum dot deep-ultraviolet light-emitting diodes with polarization-induced doping in III-nitride heterostructures. *Appl. Phys. Lett.* **104,** 021105 (2014).

5. Islam, S. M. *et al.* MBE-grown 232–270 nm deep-UV LEDs using monolayer thin binary GaN/AlN quantum heterostructures. *Appl. Phys. Lett.* **110,** 041108 (2017).

6. Carrano, J. *Chemical and biological sensor standards study* (DARPA, 2005). Available at http://go.nature.com/JNBVGN.

7. Zhang, J. P. *et al.* Pulsed atomic-layer epitaxy of ultrahigh-quality $Al_xGa_{1-x}N$ structures for deep ultraviolet emissions below 230 nm. *Appl. Phys. Lett.* **81,** 4392-4394 (2002).

8. Taniyasu, Y., Kasu, M. & Makimoto, T. An aluminium nitride light-emitting diode with a wavelength of 210 nanometres. *Nature* **441,** 325-328 (2006).

9. Orton, J. W. & Foxon, C. T. Group III nitride semiconductors for short wavelength light-emitting devices. *Rep. Prog. Phys.* **61,** 1-75 (1998).

10. Hirayama, H., Noguchi, N., Yatabe, T. & Kamata, N. 227 nm AlGaN Light-Emitting Diode with 0.15 mW Output Power Realized using a Thin Quantum Well and AlN Buffer with Reduced Threading Dislocation Density. *Appl. Phys. Expr.* **1,** 051101 (2008).

11. Hirayama, H., Noguchi, N. & Kamata, N. 222 nm Deep-Ultraviolet AlGaN Quantum Well Light-Emitting Diode with Vertical Emission Properties. *Appl. Phys. Expr.* **3,** 032102 (2010).

12. Nishida, T., Saito, H. & Kobayashi, N. Efficient and high-power AlGaN-based ultraviolet light-emitting diode grown on bulk GaN. *Appl. Phys. Lett.* **79,** 711-712 (2001).

13. Hirayama, H., Maeda, N., Fujikawa, S., Toyoda, S. & Kamata, N. Recent progress and future prospects of AlGaN-based high-efficiency deep-ultraviolet light-emitting diodes. *Jpn. J. Appl. Phys.* **53,** 100209 (2014).

14. Han, J. *et al.* AlGaN/GaN quantum well ultraviolet light emitting diodes. *Appl. Phys. Lett.* **73,** 1688-1690 (1998).

15. Pernot, C. *et al.* Development of high efficiency 255-355 nm AlGaN-based light-emitting diodes. *Phys. Status. Solidi. A.* **208,** 1594-1596 (2011).

16. Zhang, Y. *et al.* Interband tunneling for hole injection in III-nitride ultraviolet emitters. *Appl. Phys. Lett.* **106,** 141103 (2015).

17. Speck, J. S. & Rosner, S. J. The role of threading dislocations in the physical properties of GaN and its alloys. *Physica. B. Condens. Matter.* **273–274,** 24-32 (1999).

18. Kim, J. *et al.* Influence of V-pits on the efficiency droop in InGaN/GaN quantum wells. *Opt. Express* **22 Suppl 3,** A857-866 (2014).

19. Khan, A., Balakrishnan, K. & Katona, T. Ultraviolet light-emitting diodes based on group three nitrides. *Nat. Photon.* **2,** 77-84 (2008).

20. Mi, Z. *et al.* Molecular beam epitaxial growth and characterization of Al(Ga)N nanowire deep ultraviolet light emitting diodes and lasers. *J. Phys. D: Appl. Phys.* **49,** 364006 (2016).

21. Zhao, S. *et al.* Three-dimensional quantum confinement of charge carriers in self-organized AlGaN nanowires: A viable route to electrically injected deep ultraviolet lasers. *Nano Lett.* **15,** 7801-7807 (2015).

22. Painter, O. *et al.* Two-Dimensional Photonic Band-Gap Defect Mode Laser. *Science* **284,** 1819-1821 (1999).

23. Matsubara, H. *et al.* GaN Photonic-Crystal Surface-Emitting Laser at Blue-Violet Wavelengths. *Science* **319,** 445-447 (2008).

24. Sakai, M. *et al.* Random laser action in GaN nanocolumns. *Appl. Phys. Lett.* **97,** 151109 (2010).

25. Li, K. H., Liu, X., Wang, Q., Zhao, S. & Mi, Z. Ultralow-threshold electrically injected AlGaN nanowire ultraviolet lasers on Si operating at low temperature. *Nat. Nanotech.* **10,** 140-144 (2015).

26. Islam, S. M., Protasenko, V., Rouvimov, S., Xing, H. G. & Jena, D. Sub-230nm deep-UV emission from GaN quantum disks in AlN grown by a modified Stranski-Krastanov mode. *Jpn. J. Appl. Phys.* **55,** 05FF06 (2016).

27. Tchernycheva, M. *et al.* Electron confinement in strongly coupled GaN/AlN quantum wells. *Appl. Phys. Lett.* **88,** 153113 (2006).

28. Kandaswamy, P. K. *et al.* GaN/AlN short-period superlattices for intersubband optoelectronics: A systematic study of their epitaxial growth, design, and performance. *J. Appl. Phys.* **104,** 093501 (2008).

29. Xu, T. *et al.* GaN quantum dot superlattices grown by molecular beam epitaxy at high temperature. *J. Appl. Phys.* **102,** 073517 (2007).

30. Islam, S. *et al.* High efficiency deep-UV emission at 219 nm from ultrathin MBE GaN/AlN quantum heterostructures. *arXiv preprint arXiv:1704.08737* (2017).

31. Ambacher, O. *et al.* Two dimensional electron gases induced by spontaneous and piezoelectric polarization in undoped and doped AlGaN/GaN heterostructures. *J. Appl. Phys.* **87,** 334-344 (2000).

32. Peng, Z. *et al.* Polarization Induced High Al Composition AlGaN p–n Junction Grown on Silicon Substrates. *Chin. Phys. Lett.* **31,** 118102 (2014).

33. Zheng, T. C. *et al.* Improved p-type conductivity in Al-rich AlGaN using multidimensional Mg-doped superlattices. *Sci. Rep.* **6,** 21897 (2016).

34. Jena, D. *et al.* Realization of wide electron slabs by polarization bulk doping in graded III–V nitride semiconductor alloys. *Appl. Phys. Lett.* **81,** 4395-4397 (2002).

35. Li, S. *et al.* Polarization induced hole doping in graded $Al_xGa_{1-x}N$ (x = 0.7 ~ 1) layer grown by molecular beam epitaxy. *Appl. Phys. Lett.* **102,** 062108 (2013).

36. Neufeld, C. J. *et al.* Effect of doping and polarization on carrier collection in InGaN quantum well solar cells. *Appl. Phys. Lett.* **98,** 243507 (2011).

37. Sánchez-Rojas, J. L., Garrido, J. A. & Muñoz, E. Tailoring of internal fields in AlGaN/GaN and InGaN/GaN heterostructure devices. *Phys. Rev. B* **61,** 2773-2778 (2000).

38. Kozodoy, P. *et al.* Polarization-enhanced Mg doping of AlGaN/GaN superlattices. *Appl. Phys. Lett.* **75,** 2444-2446 (1999).

39. Ma, Z. & Seo, J.-H. Lattice mismatched heterojunction structures and devices made therefrom. US patent 8,866,154 (2014).

40. Dalmau, R. *et al.* Growth and Characterization of AlN and AlGaN Epitaxial Films on AlN Single Crystal Substrates. *J. Electrochem. Soc.* **158,** H530 (2011).

41. Sze, S. M. & Ng, K. K. *Physics of Semiconductor Devices*. (John Wiley & Sons, New York, 2007).

**Acknowledgement**

The work was supported by Defense Advanced Research Projects Agency (DARPA) under Grant HR0011-15-2-0002 (PM: Dr. Daniel Green).

**Author information**

Correspondence and requests for materials should be addressed to Z.M. mazq@engr.wisc.edu or J.D.A (jalbrech@egr.msu.edu) or B.M. (bmoody@hexatechinc.com)

# Supplementary Information (SI)

## I. AlN Native Oxide Growth

Aluminum compounds are reactive and oxidize rapidly since Al has a strong affinity for oxygen [1,2]. In addition, hydroxidation occurs when aluminum reacts with water vapor in the air. Thus, there exists an amorphous layer of a few nanometers thick of one or several aluminum–oxygen–hydrogen compounds such as aluminum trihydroxide ($Al(OH)_3$), aluminum oxide hydroxide ($AlOOH$ or $Al_2O_3 \cdot H_2O$), or aluminum oxide ($Al_2O_3$) [1]. It is reported [1,3] that the formation of the native hydroxides on freshly cleaned nanocrystalline AlN occurs rapidly at room temperature. The typical thickness of the aluminum oxide is 2-5 nm, as further diffusion of oxygen is prohibited by the very stable passivating layer.

This oxidation phenomenon was observed in our work. Fig. S1 shows a HRTEM image of the interface between Si and AlN/AlGaN MQW. Amorphous aluminum–oxygen–hydrogen compounds of ~7 nm thick were formed due to the exposure of the AlN/AlGaN MQW to the ambient air. In contrast to the thin and orderly interface between Si NM and GaN shown in Fig. 1c, this amorphous layer severely limits the hole carrier transport into the MQW region. As a result, a 20 nm GaN layer was grown on top of the MQW as a termination layer for the epitaxy growth, which prevents the oxidation of AlN and forms electrical contacts with the p-type Si NM.

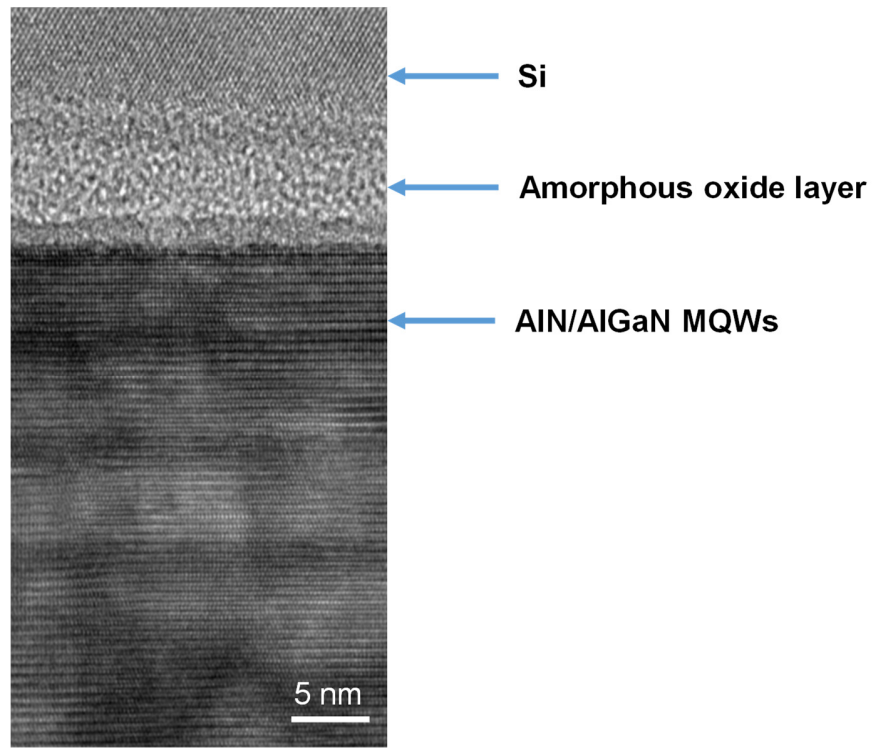

**Figure S1.** A TEM image showing native oxide between bonded Si NM and AlN/AlGaN MQWs. The native oxide was formed due to exposure of the AlGaN/GaN to air.

## II. LED Fabrication Process

Figure S2 a illustrates the LED fabrication process flow. The LED epitaxial samples were cleaned by the standard RCA method (Fig. S2 a i): sonication of the samples in acetone, isopropyl alcohol (IPA), and deionized (DI) water for 10 min to remove particles and dust from the surface, followed by immersion in piranha solution (a mixture of $H_2SO_4$ : $H_2O_2$ (4:1)) to remove metals and organic contaminants. Then the samples were immersed in SC-1 solution ($H_2O$:$H_2O_2$:$NH_4OH$ (5:1:1)) and SC-2 solution ($H_2O$:$H_2O_2$:HCl (5:1:1)) for 10 minutes per solution at room temperature to remove any remaining organic contaminants, heavy metals, and ionic contaminants. Finally, the native oxide on the wafer surface (p-GaN) was removed using diluted hydrogen fluoride (HF) (1:50 of HF:DI water) followed by a thorough rinse in deionized (DI) water before drying with nitrogen gun.

An Al$_2$O$_3$ layer of 0.5 nm was deposited by using Ultratech/Cambridge Nanotech Savannah S200 ALD system, which was also integrated with the nitrogen filled glove box. The sample was loaded in the chamber after removing native oxide with HF. It should be noted that all the processes after finishing the removal of the native oxide layer were carried out in a nitrogen environment. The chamber was pre-heated to 200°C and was pumped down to vacuum (<0.1 mTorr) immediately after loading the sample. During the ALD process, trimethylaluminum (TMA) gas and water vapor were purged for 0.015s every 5s. Five cycles of the ALD process were performed to achieve the targeted 0.5 nm thick Al$_2$O$_3$ layer as depicted in Fig. S2 a ii.

A p-type single-crystal Si nanomembrane (NM) with a doping concentration of $5 \times 10^{19}$ cm$^{-3}$ was first released from a silicon-on-insulator (SOI) substrate and was then print transferred to the AlN/AlGaN MQW structure (Fig. S2 a iii and Figure S2 b) following the procedures that we described elsewhere[4]. A thermal anneal procedure was performed under 500°C for 5 min to increase the bonding strength between the p-type Si NM and Al$_2$O$_3$ deposited epitaxial structure. Using photolithography image reversal photoresist (PR) AZ5214 for patterning, reactive ion etching (RIE) (PT790 RIE Plasma Etcher, CF$_4$: 45 sccm, O$_2$: 5 sccm, pressure: 40 mTorr, plasma power: 100 W) to etch away the Si NM and inductive coupled plasma (ICP) etcher (PT770ICP Metal Etcher, BCl$_3$/Cl$_3$/Ar: 18/10/5 sccm, pressur: 20 mTorr, plasma power: 50 W, inductor power: 500 W) to etch the p-GaN/i-AlN/AlGaN MQW, the n-type Al$_{0.82}$Ga$_{0.18}$N was exposed to form a mesa (Fig. S2 a iv and Fig. S2 c). A stack of cathode metals (Ti/Al/Ni/Au: 15/100/50/250 nm) was deposited using an e-beam evaporator and photolithography. The sample was annealed at 950°C for 30s to improve the cathode contact resistance. (Fig. S2 a v and Fig. S2 d). Another stack of anode metal (Ti/Au: 15/100 nm) was formed in the same way as the cathode metal (Fig. S2 a vi and Fig. S2 e) but without using thermal anneal. The LEDs were isolated by subsequently etching away Si NM (Fig. S2 a vii), the GaN layer, and the MQW layer n-Al$_{0.82}$Ga$_{0.18}$N layer until the AlN substrate was exposed (Fig. S2 a viii and Fig. S2 f).

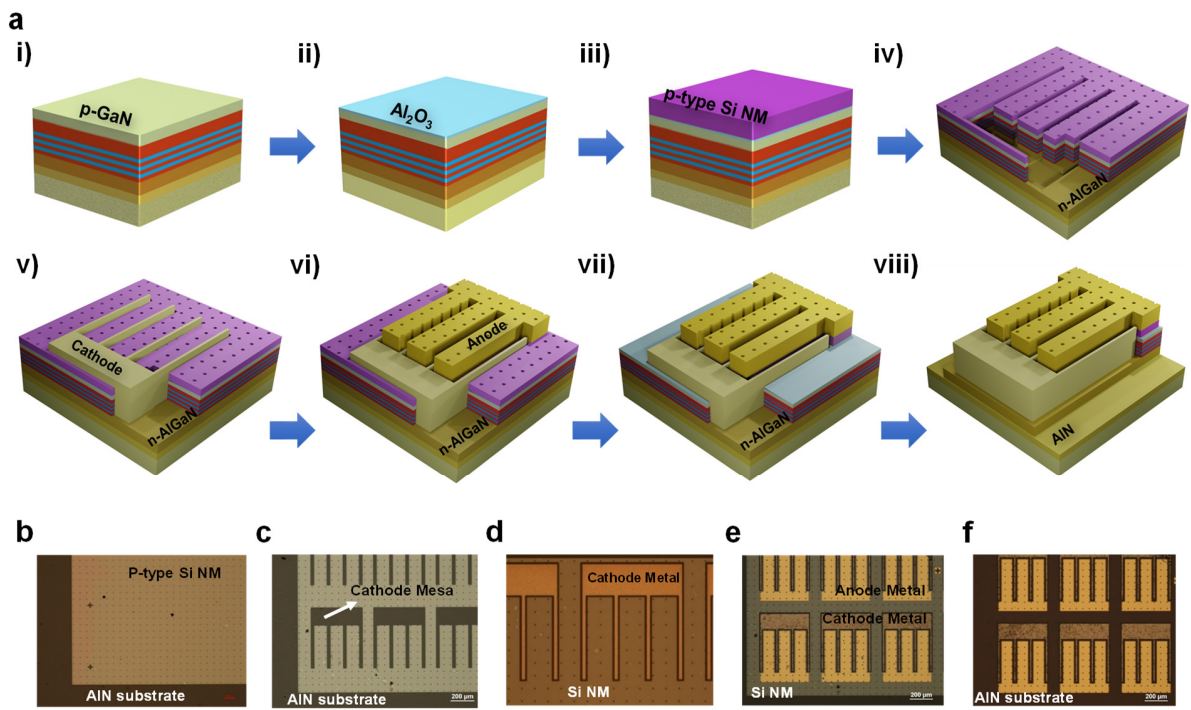

**Figure S2. Illustration of LED fabrication process and images. a,** Fabrication process flow. i) Begin with epitaxial layers and RCA cleaning. ii) Deposit 0.5 nm thick Al$_2$O$_3$ using ALD. iii) Transfer and bond Si NM. iv) n-type mesa etching to expose Al$_{0.82}$Ga$_{0.18}$N. v) Formation of cathode metal stack Ti/Al/Ni/Au: 15/100/50/250 nm. vi) Formation of anode metal stack: Ti/Au: 15/100 nm. vii) Etching away Si. viii) Etching away GaN/MQW/AlGaN to expose AlN substrate. Optical microscope images corresponding to steps in **a**: **b,** iii). **c,** iv). **d,** v). **e,** vi). **f,** a vii).

## III. XPS measurements on p-GaN surface to decide band bending before and after Al$_2$O$_3$ ALD deposition

It was reported that a Fermi level pinning state existed at 0.4 eV to 0.8 eV below the conduction band of GaN with n-type doping, due to nitrogen vacancy or gallium-dangling bond[5]. To further investigate the origin of the p-GaN surface band bending, X-ray photoelectron spectroscopy (XPS) was employed to examine the surface/interface of p-GaN before and after Al$_2$O$_3$ ALD deposition. The core level energies of Ga 3d, N 1s, Al 2p, and C 1s and valence band maximum (VBM) of p-GaN were measured with respect to the Fermi level at the surface. The measurement results are shown in Fig. S3. The 10 times-repeated scans using Al K$_\alpha$ X-ray source (hv = 1486.60 eV) were performed with 0.1 eV scan steps, 100 μm spot size, 50 eV pass energy, and 50 ms dwell time. The binding energies of the Ga 3d peak before and after Al$_2$O$_3$ deposition were 19.36 and 19.03 eV, respectively, as shown in Fig. S3 a. Thus, a 0.33 eV surface potential shift was measured on the p-GaN surface by the deposition of Al$_2$O$_3$, with respect to the as-grown p-GaN surface. Furthermore, Figure S3 b shows that the valence band maximum (VBM) of p-GaN is positioned at 1.45 eV below the Fermi level. Based on the above XPS measurement results, the band bending diagrams of p-GaN near the surface are depicted for the two scenarios (with and without Al$_2$O$_3$). Inferring from the 1.45 eV VBM in Fig. S3 b and 0.12 eV $E_F$-$E_v$, the energy bands were downward bending about -1.33 eV (Fig. S3 c left) and -1.00 eV before and after (Fig. S3 c right) the Al$_2$O$_3$ deposition, respectively. It is worth pointing out that the XPS analysis matches the band bending results obtained from C-V measurements (Fig. 2c), which reveals -0.97 eV (close enough to -1.00 eV) downward bending toward the surface of the p-GaN.

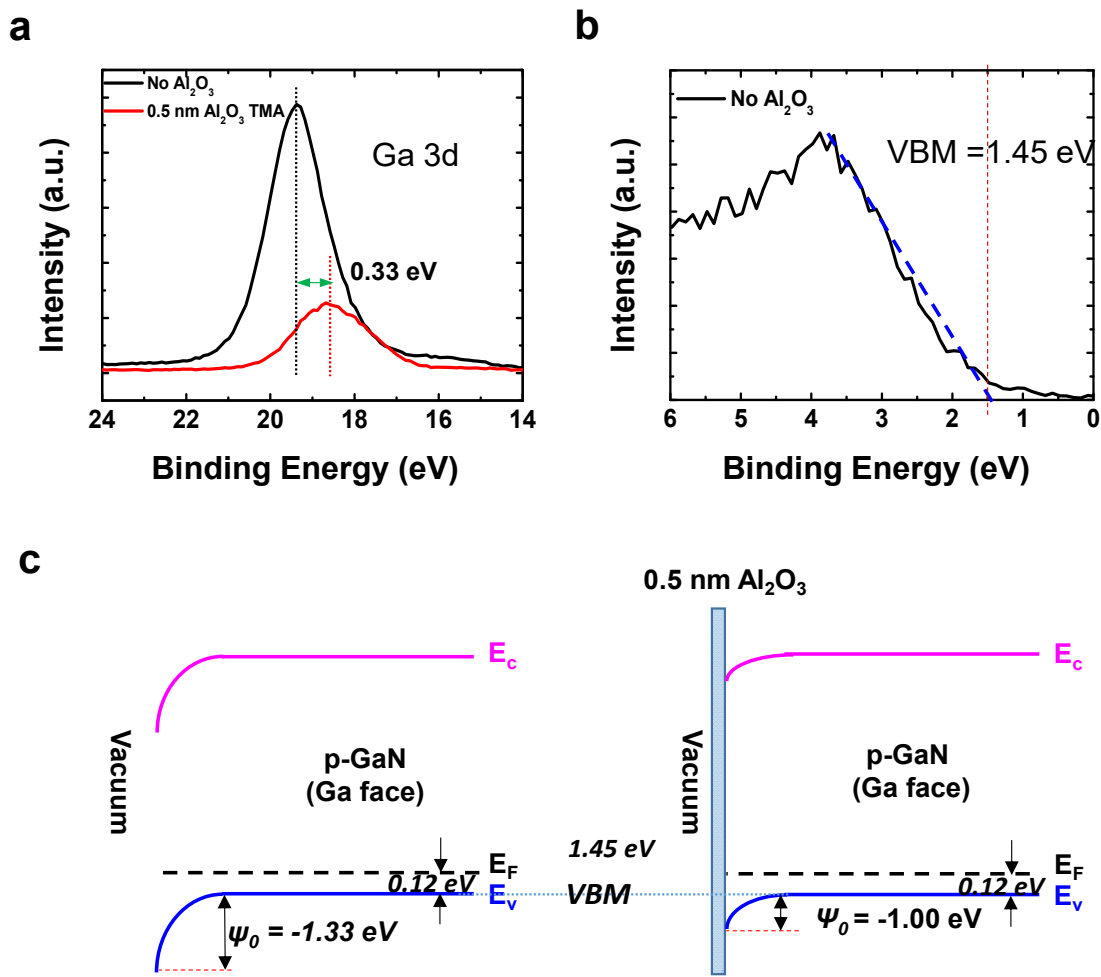

**Figure S3. XPS measurements on p-GaN surface band bending before and after $Al_2O_3$ ALD deposition. a,** Binding energies of the Ga 3d peak before and after $Al_2O_3$ deposition. **b,** VBM of p-GaN without $Al_2O_3$. **c,** Energy bands before (left) and after (right) $Al_2O_3$ deposition.

# IV. References

1    Panchula, M.L. and Ying, J.Y., Nanocrystalline Aluminum Nitride: I, Vapor-Phase Synthesis in a Forced-Flow Reactor. *J. Am. Ceram. Soc.*, *86*(7), 1114-1120 (2003).

2    Dalmau, R., Collazo, R., Mita, S. and Sitar, Z., X-ray photoelectron spectroscopy characterization of aluminum nitride surface oxides: thermal and hydrothermal evolution. *J. Electron. Mater.*, *36*(4), 414-419 (2007).

3    Dutta, I., Mitra, S. and Rabenberg, L., Oxidation of Sintered Aluminum Nitride at Near-Ambient Temperatures. *J. Am. Ceram. Soc.*, *75*(11), 3149-3153 (1992).

4    Seo, J.-H., Ling, T, Zhou, W., Gong, S., Ma, A. L., Guo. J., and Ma, Z., Fast Flexible Thin-Film Transistors With A Nanotrench Structure. *Sci. Rep.*, 6, 24771 (2016).

5    Eller, B., Yang, J. and Nemanich, R., Polarization Effects of GaN and AlGaN: Polarization Bound Charge, Band Bending, and Electronic Surface States. *J. Electron. Mater.*, *43*(12) (2014).